\documentclass[%
 reprint,
 amsmath,amssymb,
 aps,
 prl,
 longbibliography,
 lengthcheck,%
]{revtex4-1}

\usepackage{graphicx}
\usepackage{dcolumn}
\usepackage{bm}
\usepackage{hyperref}

\begin{document}

\preprint{APS/123-QED}

\title{Self-consistent spin waves in magnetized BEC 
}

\author{P. A. Andreev}%
\email{andreevpa@physics.msu.ru}
 \affiliation{Department of General Physics, Physics Faculty, Moscow State
University, Moscow, Russian Federation.}





\date{\today}

\begin{abstract}

We obtain equations of quantum hydrodynamic (QHD) for magnetized
spin-1 neutral Bose-Einstein condensate (BEC). System of QHD
equations contains the equation of magnetic moment evolution (an
analog of the Bloch equation). We account spin-spin interaction
along with the short range interaction. We consider
self-consistent field approximation of QHD equations. Starting
from QHD equation we derive the Gross-Pitaevskii equation for
magnetized BEC. We show that Gross-Pitaevskii equation exists
under condition that the magnetic moment direction is not change. Using
obtained QHD equations we study the dispersion of collective
excitation. As in electrically polarized BEC [P. A. Andreev, L. S.
Kuz'menkov, arXiv: 1201.2440], in the magnetically
polarized BEC there is second wave mode (polarization mode or spin
wave), in addition to the Bogoliubov's mode. Second wave solution appears due to
the magnetic moment evolution. The influence of magnetization on
dispersion of Bogoliubov's mode is found. We found strong
difference of dispersion properties of waves in magnetized BEC
from electrically polarized BEC.
\end{abstract}

\maketitle


\section{\label{sec:level1}I. Introduction}

Special attention at studying Bose-Einstein condensation (BEC) in ultracold atomic
vapors gives to polarized BEC ~\cite{Yi PRA
00}-~\cite{Andreev arxiv 12 02}. The term a polarized
BEC is used in literature  as for electrically ~\cite{Ni PCCP 09}, ~\cite{Carr
NJP 09} as for magnetically ~\cite{Lahaye Nature}, ~\cite{Lahaye
RPP 09} polarized BEC. In the recent years, an attention increases to
the BEC of atoms with magnetic polarization because of the BEC of
$^{52}$Cr was experimentally realized ~\cite{Lahaye Nature}. These
atoms have a large magnetic moment $6\mu_{B}$, where $\mu_{B}$ is
the Bohr magneton. There are a lot of attempts to realize
electrically polarized BEC ~\cite{Ni PCCP 09}, ~\cite{Carr NJP
09}, for this aim two nuclear molecules are  usually used.
Attention to electrically polarized BEC is particularly caused by the
fact that molecules have electrical dipole moment of order of 1
Debye and this means that the force of dipole-dipole
interaction of molecules on four orders more than the force of
spin-spin interaction of such atom as $^{52}$Cr.

Nevertheless, in the chromium atoms the quantity of spin-spin
interaction is comparable with the value of the short range
interaction. Thereby, a spin-spin interaction makes anisotropy.
Consequently, spin-spin interaction has essential contribution in
 chromium atoms static and dynamic properties.

Usually, for polarized BEC studying a generalized
Gross-Pitaevskii (GP) equation is used ~\cite{Yi PRA 00}
-~\cite{Ticknor PRL 11}. A nonlinear Schrodinger equation is usually used for quantum gases studying. One is also used for dipoles BEC description. In the last case, in GP equation a new term is added. However, this approximation includes interaction among parallel dipoles, and does not account spatial and temporal dipoles evolution. Different
approximations based on scattering process were analyzed in Ref.
~\cite{Wang NJP 08}. Particularly, where were considered condition
there can be used first Born approximation and presented generalization of GP equation
for the scattering of polarized atoms beyond first Born approximation.  In Ref. ~\cite{Wilson arxiv 11} authors developed method accounted spatially inhomogeneous distribution of dipoles directions. This method includes the potential of interaction unparallel dipoles in GP equation, and the second equation determining dipoles moment of a volume unit.
In paper ~\cite{Szirmai arxiv 11} were suggested a three fluid hydrodynamic model for description of a spin-1 Bose gas at finite temperatures, three sounds are evaluated there in a mean-field approximation.
The method of many-particle quantum hydrodynamic for electrically polarized BEC was developed in Ref. ~\cite{Andreev arxiv 12 02}. This approximation allows describe as spatial as temporal evolution of electric dipole moment. For magnetized BEC studying in this paper we use the method of
quantum hydrodynamics (QHD). This method was developed in the
recent years for various physical systems. Namely, it was quantum
plasma ~\cite{MaksimovTMP 1999}, ~\cite{Haas PRE 00}, plasma of
particles with the own magnetic moment ~\cite{MaksimovTMP
2001}-~\cite{Mahajan PRL 11}, relativistic quantum plasma
~\cite{Asenjo PP 11}, ultracold Bose and Fermi gases with nonlocal
interaction ~\cite{Andreev PRA08} and three-particle interaction
at nonzero temperatures ~\cite{Andreev arxiv ThPart}, quantum
particles with electrical polarization ~\cite{Andreev PRB 11},
particularly being in the BEC state ~\cite{Andreev arxiv 12 02},
~\cite{Andreev arxiv Pol}, the graphene electrons
~\cite{Andreev arxiv 12 01} and BEC of graphene excitons ~\cite{Andreev arxiv 12 GEBEC}. Influence of magnetic moment
evolution on the collective excitation dispersion for quantum
plasma were considered in Ref. ~\cite{Andreev VestnMSU 2007},
where were found that temporal evolution of magnetic moment leads
to existence of new branches in dispersion dependence.
More detailed analysis of spin waves in quantum plasma and methods
of these waves generation were considered in Ref. ~\cite{Andreev
arxiv MM}.

In this paper we develop a self-consistent theory for a magnetic moment evolution of the BEC of neutral spin-1 atoms. We derive the QHD equations: continuity equation, momentum balance equation (Euler equation) and magnetic moment evolution equation. The last one describe the temporal evolution of the BEC magnetization and it's influence on BEC properties. In Ref. ~\cite{Andreev arxiv 12 02} were shown that temporal evolution of collective electric polarization in BEC of particles having electric dipole moment leads to existence of new type waves in BEC - polarization waves, where supposed that BEC located in external uniform electric field. Dispersion dependence of collective excitations shows no dependence on direction of wave propagation and exhibits comparably stable excitation spectrum. Here, we study the linear dispersion in three dimensional magnetized BEC located in external magnetic field. We find the contribution of equilibrium magnetization in dispersion dependence of Bogoliubov's mode and obtain that magnetization leads to destabilization of Bogoliubov's mode. The last fact coincide to the other authors results ~\cite{Santos PRL 03}-~\cite{Fischer PRA 06R}. We show existence of magnetic moment waves (spin waves). The spin waves propagate along direction of external magnetic field.

Our paper is organized as follows. In Sec. II we describe the
used model and present the QHD equations for magnetized BEC. One
of the equations is the magnetic moment evolution equation, this
equation describe evolution of magnetic moment in space and time
due to the interaction particles magnetic moments each other and
with the external field. Corresponding GP equation is also
presented in Sec. II. Condition for GP equation derivation is
described. In Sec. III we presented the method of QHD equations
solution and the method of obtaining of the dispersion dependence.
In Sec. IV the dispersion spectrum properties of the collective
excitations are described. In Sec. V brief summary of obtained
results is presented.

\section{\label{sec:level1} II. Model}

We do not consider here the details of the QHD equation
derivation, but we present and use in the paper the QHD
equation for the neutral Bose particles with the own magnetic
moment being in the BEC state. All details for equation derivation can be found in Ref.s
~\cite{MaksimovTMP 1999}, ~\cite{MaksimovTMP 2001}, ~\cite{Andreev
PRA08}, ~\cite{Andreev PRB 11}. Here we present the Schrodinger
equation used for QHD equation derivation
$$\imath\hbar\partial_{t}\psi_{s}(R,t)=\Biggl(\Biggl(\sum_{i}\biggl(\frac{\textbf{p}^{2}_{i}}{2m_{i}}-\gamma_{i}\hat{s}^{\alpha}_{i}B^{\alpha}_{i(ext)}\biggr) $$
\begin{equation}\label{magn BEC spin gam}+\frac{1}{2}\sum_{i,j\neq i}\biggl(U_{ij}-\gamma_{i}\gamma_{j}G^{\alpha\beta}_{ij}\hat{s}^{\alpha}_{i}\hat{s}^{\beta}_{j}\biggr)\Biggr)\psi\Biggr)_{s}(R,t),\end{equation}
where we include short range and spin-spin interactions, and
action of external magnetic field on particles spin. In the Schrodinger equation
(\ref{magn BEC spin gam}) we use following designations:
$\gamma_{i}$ is the gyromagnetic ratio,
$\textbf{p}_{i}=-\imath\hbar\partial_{i}$ is the operator of
momentum, $\textbf{B}_{i(ext)}$-is the external magnetic field acting on $i$-th particle, $U_{ij}$ present the short range interaction (SRI), the Green
function of the spin-spin interaction (SSI) has form
\begin{equation}\label{magn BEC SSI Green function}G^{\alpha\beta}_{ij}=\partial^{\alpha}_{i}\partial^{\beta}_{i}(1/r_{ij})+4\pi\delta_{\alpha\beta}\delta(\textbf{r}_{ij}).\end{equation}

For spin matrixes $\hat{s}^{\alpha}_{i}$ the commutation relations
are
$$[\hat{s}^{\alpha}_{i},\hat{s}^{\beta}_{j}]=\imath\delta_{ij}\varepsilon^{\alpha\beta\gamma}\hat{s}^{\gamma}_{i}.$$

Thereby, we consider spin-1 Bose particles we present here the evident
form of the spin matrixes $\hat{s}^{\alpha}_{i}$ for particles
with spin equal to 1:

$$\begin{array}{ccc} \hat{s}_{x}=\frac{1}{\sqrt{2}}\left(\begin{array}{ccc}0&
1&
0\\
1& 0&
1\\
0& 1&
0\\
\end{array}\right),&
\hat{s}_{y}=\frac{1}{\sqrt{2}}\left(\begin{array}{ccc}0& -\imath &
0\\
\imath & 0&
-\imath \\
0& \imath &
0\\
\end{array}\right),&
\end{array}$$
$$\hat{s}_{z}=\left(\begin{array}{ccc}1&
0&
0\\
0& 0&
0\\
0& 0&
-1\\
\end{array}\right).$$

At derivation of system of QHD equations we need to write an evident
form of Hamiltonian of dipole-dipole interaction. Usually using expressions for a Hamiltonian of dipole-dipole
interaction are the same as for electric as for magnetic dipoles:
\begin{equation}\label{magn BEC widely used dd gam}H_{dd} = \frac{\delta^{\alpha\beta}-3r^{\alpha}r^{\beta}/r^{2}}{r^{3}},\end{equation}
but in this paper for SSI we used (\ref{magn BEC SSI Green function}). In Ref. ~\cite{Andreev arxiv 12 02} were shown that for electric dipoles interaction we need to use the first term in formula (\ref{magn BEC SSI Green function}).

There is well-known identity
\begin{equation}\label{magn BEC togdestvo}-\partial^{\alpha}\partial^{\beta}\frac{1}{r}= \frac{\delta^{\alpha\beta}-3r^{\alpha}r^{\beta}/r^{2}}{r^{3}}+\frac{4\pi}{3}\delta^{\alpha\beta}\delta(\textbf{r}),\end{equation}
so we can see the difference between usually using Hamiltonian and one's used in this paper (\ref{magn BEC SSI Green function}). The corrections of our choice followed from the fact that the equations obtained in the paper coincide to the Maxwell equations.

It has been shown by Breit ~\cite{spin-spin interaction}-~\cite{Lazur PRA 10}
that a Hamiltonian for spin-spin interaction, and, as a
consequence, for the interaction of magnetic moments contains a
term that is proportional to Dirac $\delta$-function
$\delta(r_{1}-r_{2})d_{1}^{\alpha}d_{2}^{\beta}$ along with (\ref{magn BEC widely used dd gam}). The coefficient
of the $\delta$-function has been refined later ~\cite{MaksimovTMP
2001} by using idea that the Hamiltonian is in accord with Maxwell's free
equations, such as $div\textbf{B}=0$. The resultant expression for
the spin-spin interaction Hamiltonian is:
\begin{equation}\label{magn BEC pot of mm int} H_{\mu\mu}=\Biggl(\frac{\delta^{\alpha\beta}-3r^{\alpha}r^{\beta}/r^{2}}{r^{3}}-\frac{8\pi}{3}\delta_{\alpha\beta}\delta(\textbf{r})\Biggr)\mu^{\alpha}_{1}\mu^{\beta}_{2}.\end{equation}
Formula (\ref{magn BEC pot of mm int}) is obtained at substitution of relation (\ref{magn BEC togdestvo}) in to the formula (\ref{magn BEC SSI Green function}).

The system of QHD equations for magnetized BEC consists of:

continuity equation
\begin{equation}\label{magn BEC cont eq}\partial_{t}n(\textbf{r},t)+\partial^{\alpha}(n(\textbf{r},t)v^{\alpha}(\textbf{r},t))=0,\end{equation}
where $n(\textbf{r},t)$ is the concentration of particles, $\textbf{v}(\textbf{r},t)$ is the velocity field. In this equation describing the whole particle number invariance, and evolution of concentration owing to particles displacement;

momentum balance equation (Euler equation)
$$mn(\textbf{r},t)(\partial_{t}+\textbf{v}\nabla)v^{\alpha}(\textbf{r},t)+\partial_{\beta}p^{\alpha\beta}(\textbf{r},t)$$
$$-\frac{\hbar^{2}}{4m}\partial^{\alpha}\triangle
n(\textbf{r},t)+\frac{\hbar^{2}}{4m}\partial^{\beta}\Biggl(\frac{\partial^{\alpha}n(\textbf{r},t)\cdot\partial^{\beta}n(\textbf{r},t)}{n(\textbf{r},t)}\Biggr)
$$
$$=\Upsilon n(\textbf{r},t)\partial^{\alpha}n(\textbf{r},t)+\frac{1}{2}\Upsilon_{2}\partial^{\alpha}\triangle n^{2}(\textbf{r},t)$$
\begin{equation}\label{magn BEC bal imp eq short}+M^{\beta}(\textbf{r},t)\partial^{\alpha}B^{\beta}(\textbf{r},t),
\end{equation}
where $\nabla$ is the the gradient operator, $\triangle$ is the
Laplace operator, $\textbf{M}(\textbf{r},t)$ is the density of
magnetic moment (magnetization), $\textbf{B}(\textbf{r},t)$ is the
magnetic field, $p^{\alpha\beta}(\textbf{r},t)$ is the kinetic
pressure describing the temperature effects and can be vanished
for the BEC,
\begin{equation}\label{magn BEC Upsilon} \Upsilon=\frac{4\pi}{3}\int
dr(r)^{3}\frac{\partial U(r)}{\partial r},
\end{equation}
and
\begin{equation}\label{magn BEC Upsilon2}\Upsilon_{2}\equiv\frac{\pi}{30}\int dr
(r)^{5}\frac{\partial U(r)}{\partial r},\end{equation} describing
the SRI influence on particles dynamic. The momentum
balance equation for the BEC with SRI up to TOIR approximation
were obtained in Ref. ~\cite{Andreev PRA08} in the absence of the
magnetization contribution. The second term in the right-hand side
of equation (\ref{magn BEC bal imp eq short}) is the term
appearing in the TOIR approximation. This term is an example of
non-local SRI, analogous terms were obtained in Ref.s
~\cite{Braaten PRA 01}, ~\cite{Rosanov PL A 02};

balance equation of magnetic moment (Bloch equation)
\begin{equation}\label{magn BEC magn evol eq}\partial_{t}M^{\alpha}(\textbf{r},t)+\nabla^{\beta}J^{\alpha\beta}_{M}(\textbf{r},t)=\frac{\gamma}{\hbar}\varepsilon^{\alpha\beta\gamma}M^{\beta}(\textbf{r},t)B^{\gamma}(\textbf{r},t),\end{equation}
where a tensor of spin current $J_{M}^{\alpha\beta}$ arises.
Vanishing by thermal motion we have
$J_{M}^{\alpha\beta}=M^{\alpha}v^{\beta}$. This equation describes
the dynamic of magnetization \emph{and} influence of external
magnetic field and inter-particle interaction on magnetization
evolution. For the first time the many-particle QHD equations for
spinning particles were derived in Ref. ~\cite{MaksimovTMP 2001}
for the case spinning charged Fermi particles. It was made in
2001.

During the derivation of QHD equation we do not consider the
scattering problem and do not use conception of scattering for
interpretation of the SRI and SSI. However, for simplicity of
comparison of our results with the results of other authors we
reduce connection of $\Upsilon$ and $\Upsilon_{2}$ with the
scattering amplitude ~\cite{Andreev PRA08}.

The first order of the interaction radius interaction constant for
dilute gases has the form $\Upsilon=-g=-4\pi\hbar^{2}a/m$, where
$g$ is the interaction constant usually used in GP equation, $a$
is the scattering length ~\cite{L.P.Pitaevskii RMP 99,Andreev
PRA08}. The value $\Upsilon_{2}$ may be expressed approximately as
$\Upsilon_{2}=-\theta a^{2}\Upsilon/8$ ~\cite{Andreev PRA08},
where $\theta$ is a constant positive value about 1, which depends
on the interatomic interaction potential. Finally, $\Upsilon_{2}$
takes the form $\Upsilon_{2}=-\pi\theta\hbar^{2}a^{3}/2m$.

We consider the three-dimensional (3D) system of Bose particles.
It allows us to introduce in equations (\ref{magn BEC bal imp eq
short}) and (\ref{magn BEC magn evol eq}) the magnetic field
caused by particles magnetic moments. Spin-spin interaction in
these equations at derivation appears in the integral form (see
for example ~\cite{MaksimovTMP 2001}). In self-consistent
approximation in 3D particles system we can represent interaction
via magnetic field which has form
$$B^{\alpha}(\textbf{r},t)=\int d\textbf{r}' G^{\alpha\beta}(\mid\textbf{r}-\textbf{r}'\mid)M^{\beta}(\textbf{r}',t),$$
which satisfies to the Maxwell equation:
\begin{equation}\label{magn BEC Maxwell}curl \textbf{B}(\textbf{r},t)=4\pi curl \textbf{M}(\textbf{r},t).\end{equation}

From equations (\ref{magn BEC cont eq})-(\ref{magn BEC magn evol
eq}) we can see that short range interaction, the terms
proportional to $\Upsilon$ and $\Upsilon_{2}$, gives
contribution in the momentum balance equation only.

In the absence of magnetic moment direction evolution
$M^{\alpha}(\textbf{r},t)=n(\textbf{r},t)\mu^{\alpha}(\textbf{r},t)$,
$\mu^{\alpha}(\textbf{r},t)=const$ (supposing that magnetization
change because of the concentration changing only), we can derive
the GP equation for spinning neutral particles in the BEC state.
We notice that for the $\mu^{\alpha}(\textbf{r},t)=const$  the
magnetic moment evolution equation does not exist (It does not need). In this case
the magnetic moment evolution reduces to the concentration
evolution $\textbf{M}(\textbf{r},t)\sim n(\textbf{r},t)$.
Particles dynamic is described by two equations: the continuity
equation (\ref{magn BEC cont eq}) and simplification of the Euler
equation (\ref{magn BEC bal imp eq short}).

The method of GP equation derivation from QHD equation is
described in Ref.s ~\cite{Andreev PRA08} and ~\cite{Andreev PRB
11}. Resulting GP equation, in the absence of terms appeared up to TOIR approximation, has form
$$\imath\hbar\partial_{t}\Phi(\textbf{r},t)=\Biggl(-\frac{\hbar^{2}}{2m}\nabla^{2}+g\mid\Phi(\textbf{r},t)\mid^{2}$$
\begin{equation}\label{magn BEC nlse int general} +\mu_{0}^{\beta}\mu_{0}^{\gamma}\int
d\textbf{r}'G^{\beta\gamma}(\textbf{r},\textbf{r}')|\Phi(\textbf{r}',t)|^{2}\Biggr)\Phi(\textbf{r},t),\end{equation}
where $G^{\alpha\beta}$ the Green function of SSI (\ref{magn BEC SSI Green function}), using evident form of the Green function we can rewrite GP equation
$$\imath\hbar\partial_{t}\Phi(\textbf{r},t)=\Biggl(-\frac{\hbar^{2}}{2m}\nabla^{2}+g\mid\Phi(\textbf{r},t)\mid^{2}$$
$$+\mu_{0}^{2}\int
d\textbf{r}'\frac{1-3\cos^{2}\theta
'}{|\textbf{r}-\textbf{r}'|^{3}}|\Phi(\textbf{r}',t)|^{2}$$
\begin{equation}\label{magn BEC nlse int final} -\frac{8\pi}{3}\mu_{0}^{2}|\Phi(\textbf{r},t)|^{2}\Biggr)\Phi(\textbf{r},t).\end{equation}
This equation has difference from usually used for magnetized BEC
~\cite{Goral PRA 00}, ~\cite{Santos PRL 00}, ~\cite{Wilson PRL
10}, ~\cite{Ticknor PRL 11}. This difference appear because of
second term in (\ref{magn BEC SSI Green function}) proportional to
delta function. The last term in equation (\ref{magn BEC nlse int
final}) appear because of the second term in Green function of SSI
(\ref{magn BEC pot of mm int}).

At low temperature in the external magnetic field the atoms
magnetic moments directed along the external field, because of
absence disordered factors as a thermal motion. However, the
perturbation of dipoles direction might propagate there. These
perturbations can exhibits as waves. In this case we have deal
with the spin waves. It is very spread phenomenon in the system of
particles with magnetic moment. For example we can admit that spin
waves there are in ferromagnetic ~\cite{Kittel Introduction 1999},
but there magnetic moments bound with the unit of crystalline
lattice. In gases atoms can move through the system. Thus, in
gases there is one more mechanism of magnetization changing,
compare with ferromagnetic. On the other hand, the Bose-Einstein
condensation of spin wave quanta has been also studied
~\cite{Dzyapko PTRS A}.

Different generalization of GP equation has been used in
literature. One of them is the spinor nonlinear Schrodinger
equation ~\cite{Szankowski PRL 10}. This equation is equivalent to
the three hydrodynamic equations ~\cite{MaksimovTMP 2001}, these
are continuity equation, Euler equation and magnetic moment
balance equation. Whereas, usual scalar GP equation equivalents to
continuity and Euler equations only. Thus, the spinor
generalization of GP equation also described the magnetic moment
evolution. Consequently, this equation contains information about
spin waves considered in our paper. At derivation of the
hydrodynamic equations from the spinor nonlinear Schrodinger
equation the SSI appears in self-consistent field approximation,
without two-particle quantum correlation. In fact,  quantum
spin-spin correlations might be included in GP equation if they
have a simple enough form.

In the case of spin-0 particles or if we do not account magnetic
moments motion there is the collective excitation in the BEC, it
is the Bogoliubov mode. In Ref. ~\cite{Andreev arxiv 12 02} were
shown that electric dipoles evolution leads to the existence of new
wave mode, it is the wave of electrical polarization. In this
paper, we show the existence of the spin waves (magnetic moment
waves) in magnetized BEC. Below we consider wave dispersion
calculations based on above described model.

\section{\label{sec:level1} III. Method of solution}

We can analyze the linear dynamics of elemental excitations in the
magnetized 3D BEC using the QHD equations (\ref{magn
BEC cont eq}), (\ref{magn BEC bal imp eq short}), (\ref{magn BEC
magn evol eq}) and (\ref{magn BEC Maxwell}). Let's assume the
system is placed in an external magnetic field
$\textbf{B}_{0}=B_{0}\textbf{e}_{z}$. The values of equilibrium
concentration $n_{0}$ and equilibrium magnetization
$\textbf{M}_{0}\sim\textbf{B}_{0}$ for the system in an
equilibrium state are constant and uniform and its velocity field
$v^{\alpha}(\textbf{r},t)$ values equal to zero.

We consider the small perturbation of equilibrium state
$n=n_{0}+\delta n$, $v^{\alpha}=0+v^{\alpha}$ and
$M^{\alpha}=M_{0}^{\alpha}+\delta M^{\alpha}$. Substituting these
relations into system of equations (\ref{magn BEC cont eq}),
(\ref{magn BEC bal imp eq short}), (\ref{magn BEC magn evol eq})
and (\ref{magn BEC Maxwell}) \textit{and} neglecting nonlinear
terms, we obtain a system of linear homogeneous equations in
partial derivatives with constant coefficients. Passing to the
following representation for small perturbations $\delta f$
$$\delta f =f(\omega, \textbf{k}) exp(-\imath\omega t+\imath \textbf{k}\textbf{r}) $$
one yields the homogeneous system of algebraic equations. The magnetic
field strength is assumed to have a nonzero value. Expressing all
the quantities entering the system of equations in terms of the
magnetic field, we come to the equation
$\Lambda^{\alpha\beta}(\omega, \textbf{k})\cdot B^{\beta}(\omega,
\textbf{k})=0,$ where $\Lambda^{\alpha\beta}(\omega, \textbf{k})$
is the dispersion matrix. The evident form of the dispersion
matrix $\Lambda^{\alpha\beta}(\omega, \textbf{k})$ is presented in
the Appendix. In this case, the dispersion equation is
$det\widehat{\Lambda}(\omega, \textbf{k})=0.$ Solving this
equation with respect to $\omega^{2}$ we obtain following results.

\section{\label{sec:level1}IV. Elementary excitations in the polarized BEC}

The dispersion characteristic for EE in BEC can be expressed in
the form of
\begin{equation}\label{magn BEC disp dep Bogoliubov branch}\omega^{2}=\biggl(\frac{\hbar^{2}}{4m^{2}}+\frac{n_{0}\Upsilon_{2}}{m}\biggr)k^{4}-\frac{n_{0}}{m}(\Upsilon+4\pi\mu_{0}^{2})k^{2},\end{equation}
the equilibrium magnetization $\mu_{0}$ caused by external
magnetic field $B_{0}$ and has form $\mu_{0}=\chi B_{0}/n_{0}$,
where $\chi$ is the ratio between equilibrium magnetic
susceptibility and magnetic permeability. In the case when all
magnetic dipoles directed parallel to external field (in equilibrium state) we have
$\mu_{0}=\gamma$. From formula (\ref{magn BEC disp dep Bogoliubov
branch}) we can see that for the repulsive SRI ($\Upsilon<0$) at
long wave length limit magnetization lead to instability of the
Bogoliubov's mode because the frequency square become negative:
$$\omega^{2}=\frac{n_{0}}{m}(\mid\Upsilon\mid-4\pi\mu_{0}^{2})k^{2}<0$$
at condition $4\pi\mu_{0}^{2}>\mid\Upsilon\mid$.

We obtained three solutions for spin waves. They frequencies are
$$\omega=\Omega ,$$
\begin{equation}\label{magn BEC disp dep spin wave branch}\begin{array}{ccc}   & \omega=\Omega (1\pm 4\pi\chi),& \end{array} \end{equation}
where
$\Omega=\gamma B_{0}/\hbar$ is the cyclotron frequency. It is the frequency of the single magnetic moment precession in the external uniform magnetic field.
For the case when all magnetic dipoles directed parallel to external field fully magnetized BEC, which can be considered as ferromagnetic state, we have
\begin{equation}\omega=\Omega (1\pm 4\pi\chi)=\left(\begin{array}{ccc}\Omega (1+8\pi)&
&
\\
& &
\\
\Omega& &
\\
\end{array}\right).\end{equation}
Thus, we have only two solution in this case because one of solutions $\omega=\Omega (1\pm 4\pi\chi)$ equal to $\Omega$.

\section{\label{sec:level1} V. Conclusion}

We researched dispersion properties of magnetized 3D BEC. We obtained
the contribution of equilibrium magnetization in the Bogoliubov's
mode dispersion dependence. We found that time evolution of
magnetic moments lead to existence of new type of waves in
magnetized BEC, it is spin waves. We found that SRI has no
influence of spin wave dispersion. At comparison dispersion
properties magnetized BEC with the electrically polarized BEC we
got large difference between them.

We studied dispersion of collective excitations by means of QHD equations derived in this paper for magnetized BEC.

We have found two types of branches of dispersion curve. These are Bogoliubov's mode and spin waves.

In correspondence with the other authors we obtained that
magnetization leads to destabilizing of Bogoliubov's mode. For
attractive SRI at small wave vector we have
$\omega^{2}=-n_{0}\mid\Upsilon\mid/m\cdot k^{2}$
$=-4\pi\hbar^{2}\mid a\mid n_{0}/m^{2}\cdot k^{2}$, this spectrum
is unstable. Addition of magnetization leads to instability
growing
$\omega^{2}=-(n_{0}/m)(\mid\Upsilon\mid+4\pi\mu_{0}^{2})k^{2}$.
For repulsive SRI $\Upsilon=-\mid\Upsilon\mid$ and in the absence
of the magnetization we have stable sound wave
$\omega^{2}=n_{0}\mid\Upsilon\mid/m\cdot k^{2}$, but magnetization
disturbs stability
$\omega^{2}=(n_{0}/m)(\mid\Upsilon\mid-4\pi\mu_{0}^{2})k^{2}$.
This spectrum remains stable in two cases. The first one while
magnetization is small compare with the constant of SRI
$\Upsilon$. The second case corresponds to the large constant of
SRI $\Upsilon$ which can be reached by the Feshbach resonance
~\cite{Chin RMP 10}-~\cite{Bloch RMP 08}.

In general case in magnetized BEC exist three spin waves with
frequencies: $\omega=\Omega$ and $\omega=\Omega(1\pm 4\pi\chi)$.
But, for fully magnetized ferromagnetic BEC $\Omega(1-4\pi\chi)$
equal to $\Omega$, so we have only two solutions $\omega=\Omega$
and $\omega=\Omega(1+8\pi)$. As any physical system of particles
with magnetic moments, especially ferromagnets, the magnetized BEC
shows existence of the spin waves. In this paper we consider the
macroscopic spin dynamic described by QHD, but there are a lot of
well-known method's based on second quantization
~\cite{Landau9}-~\cite{Hofmann PRB 11}. These methods are close to
the Bese-Habbard model and can be used together for studying of
microscopic influence of magnetic moment dynamic on quantum gases
properties.

\section{\label{sec:level1} Acknowledgments}

The author thanks Professor L. S. Kuz'menkov for fruitful
discussions.

\section{\label{sec:level1} Appendix: Evident form of dispersion matrix}

We describe $\Lambda^{\alpha\beta}(\omega, \textbf{k})$ by
presenting of the evident form of each component of the matrix
$$\Lambda^{xx}(\omega, \textbf{k})=4\pi\omega k_{z}n_{0}\frac{\gamma\mu_{0}}{\hbar}\frac{1}{\omega^{2}-\frac{\gamma^{2}B_{0}^{2}}{\hbar^{2}}};$$
$$\Lambda^{xy}(\omega, \textbf{k})=\imath k_{z}+4\pi\imath k_{z}n_{0}\frac{\gamma B_{0}}{\hbar}\frac{\gamma\mu_{0}}{\hbar}\frac{1}{\omega^{2}-\frac{\gamma^{2}B_{0}^{2}}{\hbar^{2}}};$$
$$\Lambda^{xz}(\omega, \textbf{k})=-\imath k_{y}+4\pi\imath k_{y}n_{0}\mu_{0}\Xi ;$$
$$\Lambda^{yx}(\omega, \textbf{k})=(\Lambda^{xy}(\omega, \textbf{k}))^{*};$$
$$\Lambda^{yy}(\omega, \textbf{k})=\Lambda^{xx}(\omega, \textbf{k});$$
$$\Lambda^{yz}(\omega, \textbf{k})=\imath k_{x}-4\pi\imath k_{x}n_{0}\mu_{0}\Xi ;$$
$$\Lambda^{zx}(\omega, \textbf{k})=\imath k_{y}$$
$$+4\pi n_{0}\frac{\gamma\mu_{0}}{\hbar}\frac{1}{\omega^{2}-\frac{\gamma^{2}B_{0}^{2}}{\hbar^{2}}}\biggl(-\omega k_{x}+\imath k_{y}\frac{\gamma B_{0}}{\hbar}\biggr);$$
$$\Lambda^{zy}(\omega, \textbf{k})=-\imath k_{x}$$
$$+4\pi n_{0}\frac{\gamma\mu_{0}}{\hbar}\frac{1}{\omega^{2}-\frac{\gamma^{2}B_{0}^{2}}{\hbar^{2}}}\biggl(-\omega k_{y}-\imath k_{x}\frac{\gamma B_{0}}{\hbar}\biggr);$$
$$\Lambda^{zz}(\omega, \textbf{k})=0,$$
where
$$\Xi=\frac{-\mu_{0}k^{2}}{m\omega^{2}-\frac{\hbar^{2}}{4m}k^{4}+(\Upsilon+4\pi\mu_{0}^{2})n_{0}k^{2}-n_{0}\Upsilon_{2}k^{4}}.$$
Quantity $\Xi$ contains whole information about SRI.


\end{document}